# Ion Trap in a Semiconductor Chip

D. Stick[1], W. K. Hensinger[1], S. Olmschenk[1], M. J. Madsen[1], K. Schwab[2], and C. Monroe[1]

*[1] FOCUS Center and University of Michigan Department of Physics, Ann Arbor, MI 48109*
*[2] Laboratory for Physical Sciences, College Park, MD 20740*

The electromagnetic manipulation of isolated atoms has led to many advances in physics, from laser cooling[1] and Bose-Einstein condensation of cold gases[2] to the precise quantum control of individual atomic ions[3]. Work on miniaturizing electromagnetic traps to the micrometer scale promises even higher levels of control and reliability[4]. Compared with 'chip traps' for confining neutral atoms[5,6,7], ion traps with similar dimensions and power dissipation offer much higher confinement forces and allow unparalleled control at the single-atom level. Moreover, ion microtraps are of great interest in the development of miniature mass spectrometer arrays[8], compact atomic clocks[9], and most notably, large scale quantum information processors[10,11]. Here we report the operation of a micrometer-scale ion trap, fabricated on a monolithic chip using semiconductor micro-electromechanical systems (MEMS) technology. We confine, laser cool, and measure heating of a single $^{111}Cd^+$ ion in an integrated radiofrequency trap etched from a doped gallium arsenide (GaAs) heterostructure.

Current ion trap research is largely driven by the quest to build a quantum information processor[12], where quantum bits (qubits) of information are stored in individual atomic ions and connected through a common interaction with a phonon[3,13] or photon[14,15] field. The fundamental experimental requirements for quantum processing have all been met with ion traps, including demonstrations of multi-qubit quantum gates and small algorithms[16,17,18]. Effort in this area is now focused on the scaling of ion traps to host much larger numbers of qubits, perhaps by shuttling individual atoms through a complex maze of ion trap electrodes[10,11]. The natural host for such a scalable system is an integrated ion trap chip. We confine single $^{111}Cd^+$ qubit ions in a radiofrequency linear ion trap[3,19] on a chip by applying a combination of static and oscillating electric potentials to integrated electrodes[20]. The electrodes are lithographically patterned from a monolithic semiconductor substrate, eliminating the need for manual assembly and alignment of individual electrodes. The scaling of this structure to hundreds or thousands of electrodes thus seems possible with existing semiconductor fabrication technology.

Candidate linear ion trap geometries amenable to microfabrication include (i) symmetric high-aspect-ratio multilayer structures with electrodes surrounding the ions[20], and (ii) asymmetric planar structures with the ions residing above a planar array of electrodes[21]. The symmetric geometry demonstrated here may be more difficult to fabricate than the asymmetric geometry, but it is deeper, has better optical access, and is less sensitive to electric field noise from correlated potentials on the electrodes (e.g., applied voltage noise or radiofrequency thermal fields common to the electrodes[8,9]). A symmetric ion trap fabricated from silicon electrodes has been demonstrated[22], requiring manual assembly and alignment of separated electrode sections. Here we report an integrated ion trap fashioned from a monolithic microchip that does not require assembly and is therefore suitable for miniaturization and scaling.

The trap is fabricated from four alternating layers of aluminum gallium arsenide (AlGaAs) and gallium arsenide (GaAs) epitaxially grown on a GaAs substrate, as described in the Methods section and illustrated in Figs. 1 and 2. The two GaAs layers (thickness 2.3 μm) are highly doped (~$3\times10^{18}$ $e$/cm$^3$) and formed into cantilevered electrodes surrounding the free-space trap region. A through-hole is etched in the substrate allowing clear optical access. The electrodes are electrically isolated from each other and from the doped GaAs substrate by the interleaved AlGaAs layers (thickness $h = 4$ μm). These insulating layers are undercut ~15 μm from the tips of the GaAs cantilevers to shield the trapped ion from stray charge on the exposed insulator. The electrodes are segmented along the axis of the linear trap, as shown in Fig. 1d. Each of the four segments has an axial width of $w = 130$ μm and is separated from adjacent segments by a 25 μm gap. The tip-to-tip separation between opposing cantilevers in the plane of the chip is $s = 60$ μm. A radiofrequency potential is applied to all axial segments of the top GaAs cantilevers on one side of the trap and bottom cantilevers on the opposite side. Static potentials are applied to the other cantilevers, which are held near radiofrequency ground with on-board filters. Ions can be trapped in one of two zones with appropriate static potentials applied to the four segments. Each of the local trap zones is primarily controlled by three adjacent segments: two endcap segments surrounding a center segment nearest to the ion. Mechanical resonances of the cantilevers are expected to occur in the 1-10 MHz range[20], with quality factors expected to be of order $10^3$.

Ovens containing cadmium oxide are heated to produce a vapor of cadmium in the trapping region with an estimated partial pressure of ~$10^{-11}$ torr. We photoionize the cadmium atoms by directing laser pulses (~100 fs pulse duration at 80 MHz repetition rate) into the trapping region that are tuned near the neutral cadmium $^1S_0 \rightarrow {}^1P_1$ transition at 228.5 nm with about 1 mW of average power focused down to a ~20 μm waist. We selectively load and Doppler laser-cool $^{111}Cd^+$ isotopes by adding a continuous-wave laser red-tuned within one natural linewidth of the $^{111}Cd^+$ $^2S_{1/2} \rightarrow {}^2P_{3/2}$ transition near 214.5 nm (all other Cd$^+$ isotopes are Doppler heated). The

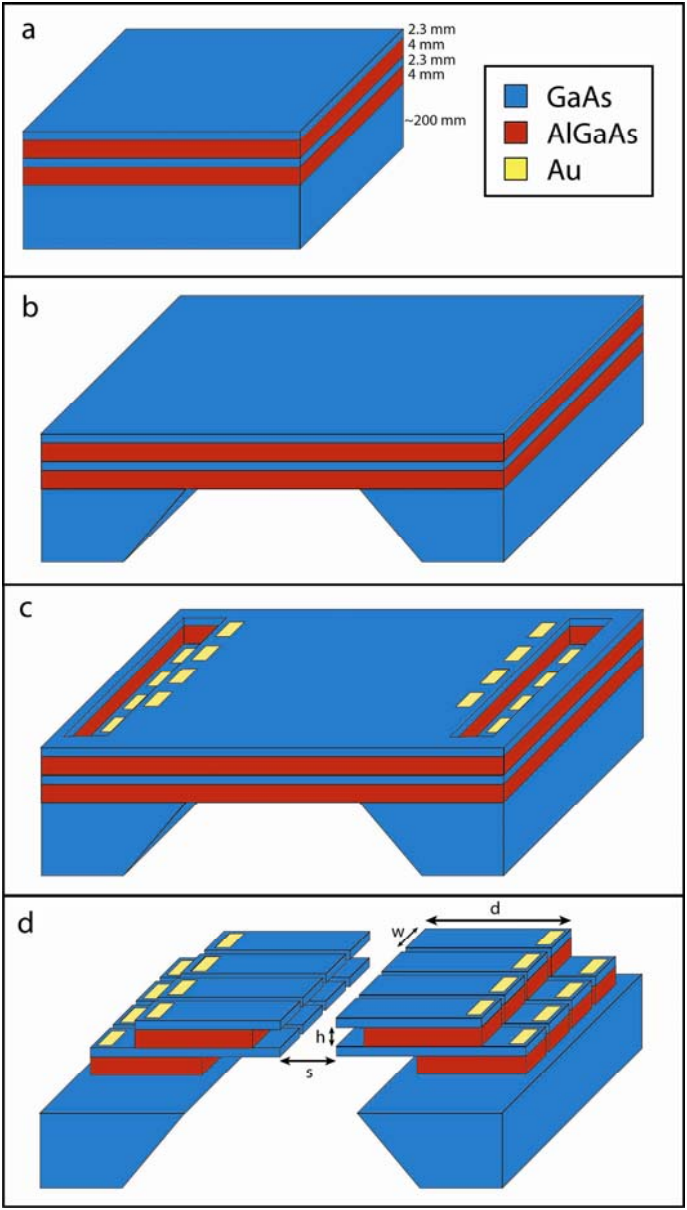

**Figure 1**: Fabrication process for a semiconductor ion trap. **(a)** The structure grown by molecular beam epitaxy consists of alternating GaAs/AlGaAs membrane layers on a GaAs substrate. **(b)** Backside etch removes substrate material for clear optical access through the chip. **(c)** Inductively-coupled plasma etch through membrane creates access to submerged GaAs layers, and gold/germanium bond pads are deposited for electrical contacts to the trap electrodes. **(d)** A further inductively-coupled plasma etch through the membrane defines and isolates the cantilevered electrodes, and a hydrofluoric acid etch undercuts the AlGaAs insulator material between the electrodes.

Doppler-cooling laser has up to 1 mW of power focused down to a ~15 μm waist. With both beams aligned, a single $^{111}$Cd$^+$ ion can be loaded after a few seconds, after which time the photoionization laser is blocked. The ion is imaged with a charge-coupled-device camera to a nearly diffraction-limited spot with f/2.1 optics, where f is the focal length, as displayed in Fig. 3. Storage lifetimes in excess of 1 h are observed, and a histogram of many loads shows an exponentially-distributed confinement time with a mean lifetime of 10 min when the ion is continuously Doppler-cooled.

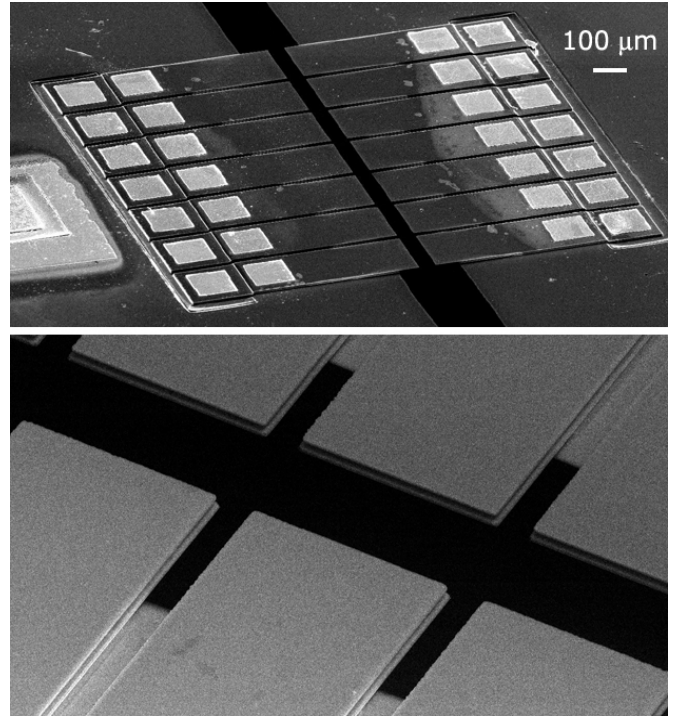

**Figure 2**: Scanning electron microscope image of a monolithic GaAs semiconductor linear ion trap. TOP: Ion trap chip with seven axial segments (28 electrodes) cantilevered over a rectangular through-hole (black). The 28 gold bonding pads are visible as bright squares, along with a single bond pad at the left connecting to the substrate beneath. In the experiment, we trap ions in a similar structure with four segments instead of seven. The tip-to-tip separation of electrodes across the gap is $s = 60$ μm. BOTTOM: Closeup of a single ion trap segment, clearly showing the upper and lower GaAs layers separated by $h = 4$ μm. The microscope was a JEOL 6500.

We directly measure the frequency of small oscillations of the trapped ion by applying a weak, variable frequency potential to one of the electrodes and observing changes in the ion fluorescence owing to the resonant force while it is continuously laser-cooled[23]. For an applied radiofrequency potential amplitude of $V_0 = 8.0$ V at a drive frequency of $\Omega_T/2\pi$ = 15.9 MHz (see Methods section), and static potentials of 1.00 V on the endcap electrodes and −0.33 V on the center electrodes, we measure the axial secular frequency to be $\omega_z/2\pi$ = 1.0 MHz. The measured transverse secular frequencies are $\omega_x/2\pi$ = 3.3 MHz and $\omega_y/2\pi$ = 4.3 MHz, indicating a radiofrequency trap stability factor[19] of $q = 0.62$. These measurements are consistent with a 3-dimensional numerical simulation of the trapping potential, which further indicates that one of the transverse principal axes of the trap is rotated ~40° out of the plane of the chip[20].

Microscale ion traps are expected to be particularly sensitive to noisy potentials from the electrodes[24,25]. Uncontrolled static offset electric fields from accumulated charge on insulating surfaces or contact potentials can give rise to radiofrequency micromotion[19] and they can even destabilize the trap. We suppress micromotion along the direction of the Doppler cooling beam by applying static offset potentials to electrodes that minimize both the broadening of the atomic fluorescence spectrum (half-width of ~50 MHz, to be compared with the natural half-width of 30 MHz) and the time-correlation of the atomic fluorescence with the radiofrequency trap drive frequency[26]. We measure heating of the secular motion of the trapped ion by performing optical stimulated Raman spectroscopy on the hyperfine qubit levels of the ion[24,25]. As described in the methods section and shown in Fig 4a, we extract a heating rate along the axial dimension of $(1.0\pm0.5)\times10^6$ quanta s$^{-1}$, at an axial trap frequency of 0.9 MHz. From this, we infer a resonant electric field noise level of about $2.0\times10^{-8}$ (V/m)$^2$/Hz (ref. 24). This is in the range of what might be expected on the basis of previous Cd$^+$ ion trap structures[25], assuming a $1/d^4$ scaling of the noise field with distance $d$ between the ion and the nearest electrode[24], and is roughly three orders of magnitude larger than the expected level of thermal electric field noise from the resistive electrodes. The source of the observed heating is unknown, but may be related to fluctuating 'patch' potentials on the electrode surfaces[24]. The interaction between the ion and driven mechanical motion of the cantilevers may also play a role, and this interesting interface between atomic and solid-state systems will be investigated in the future[27,28].

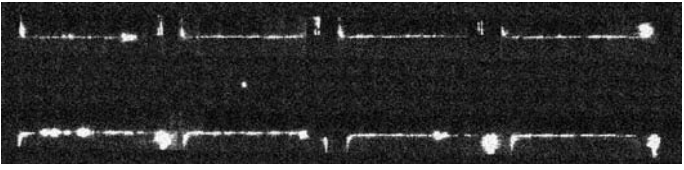

**Figure 3**. An image of a single trapped Cd$^+$ ion along a view perpendicular to the chip plane after ~1 s of integration. The ion fluoresces from applied laser radiation directed through the chip at a 45° angle and nearly resonant with the Cd$^+$ $^2S_{1/2} - ^2P_{3/2}$ electronic transition at a wavelength of 214.5 nm. The fluorescence is imaged onto a charge-coupled-device camera with an $f/2.1$ objective lens, resulting in a near-diffraction-limited spot with ~1 μm resolution at the ion. The profile of the electrodes is also clearly visible as scattered radiation from a deliberately misaligned laser that strikes the trap electrodes. The vertical gap between the top and bottom set of electrodes is $s = 60$ μm.

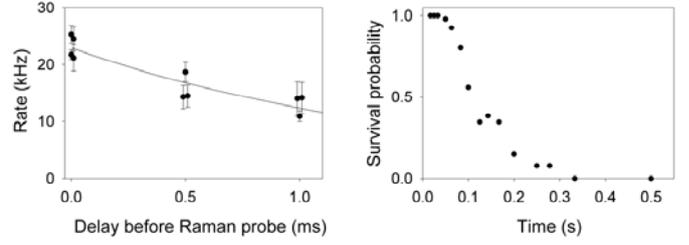

**Figure 4**. Heating rate measurements of a single Cd$^+$ ion in the microtrap. **(a)** Measurement of the motion-sensitive stimulated Raman transition rate between hyperfine states in $^{111}$Cd$^+$ vs. the delay time τ=0, 0.5, and 1.0 ms before the Raman probe (overlapping data are separated slightly for clarity). The curve is an exponential fit to the data, with the decay constant related to the heating rate. Given the Lamb-Dicke parameter of 0.018 and trap frequency of 0.9 MHz, this implies a heating rate of $d\overline{n}(\tau)/d\tau = (1.0\pm0.5)\times10^6\ s^{-1}$, where $\overline{n}$ is the average harmonic vibrational index. The error in the heating rate is dominated by systematic uncertainties in relating the Raman transition rate to the heating rate, in addition to the ± 1σ error bars shown in the figure, which are due to the uncertainty in the fit of the initial Raman transition data. **(b)** A histogram of the observed survival probability of a single ion in the trap after various times in the dark without Doppler cooling (500 events total). Errors are calculated based on the underlying Bernoulli sampling process. The clear knee in the data indicates that the ion is boiled out of the trap after about 0.1 s.

To reliably load, store, and shuttle ions, a microscale trap must have sufficient depth, defined as the amount of energy needed for an ion to escape. Numerical simulations indicate that the trap depth is approximately $\Delta \sim 0.08$ eV for the above conditions, limited in a direction inclined by ~37° out of the plane of the chip. This relatively shallow depth, of order room temperature (0.025 eV), corroborates further observations of the chip trap behavior: the mean storage time of 10 minutes is consistent with the expected time between elastic collisions[27] with the room-temperature background gas (primarily Cd), and we were never able to load two ions in the trap simultaneously. Without continuous Doppler cooling, the ion is observed to boil out of the trap within the dark time, where the delay time τ = 0.1 s (Fig. 4b), implying an average heating rate $\Delta/\tau$ that is approximately 100 times higher than the heating rate measured near the bottom of the trap reported above[29]. All of these observations contrast sharply with the behavior of larger (mm scale) Cd$^+$ trap structures in our laboratory with depths > 1 eV and similar background pressures, where the storage lifetime is typically measured in days (even without laser-cooling), and multiple ions are easily loaded.

The transverse depth of a linear radiofrequency ion trap scales as $D = \sigma q e V_0/8$, where q is the stability factor, $e$ is the charge of the ion and $\sigma \leq 1$ is a geometrical shape factor. If we assume that the radiofrequency potential amplitude is limited by $V_0 \propto E_{max}\,l$, with $E_{max}$ the maximum electric field (given by electrical breakdown, field emission, or other limits), and $l$ is the limiting dimension of the trap electrodes, then the trap depth scales as $D \propto l$ (ref. 19). In this scaling law, we assume $q$ is fixed, and all dimensions are scaled isotropically ($\sigma$ = constant). However, any reduction in size will primarily be in the plane of the chip (shrinking dimensions $s$ and $w$ in Fig. 1d, but not the layer separation $h$). In this case, the depth $D \approx \sigma(s) q e E_{max} h/8$ should actually improve as $s$ becomes smaller than the geometrical shape factor $\sigma(s) \propto s^{-0.44}$ (for aspect ratios $1 < s/h < 20$)[20].

Another concern in the operation of ion microtraps is radiofrequency power dissipation, which limits the applied $V_0$ and $\Omega_T$. In general, the power dissipated in a radiofrequency ion trap is given by $P_D = V_0^2 C\Omega_T/(2Q)$, where the quality factor $Q$ describes the radiofrequency losses in the trap structure and is given by $1/Q = R_SC\Omega_T + tan\delta$. Here, $C$ is the net capacitance and $R_S$ the net series resistance of the radiofrequency electrodes, and $tan\delta$ is the loss tangent of the insulating layer. In the experiment, we measure $Q \sim 55$ from the radiofrequency resonance shape. This is consistent with a direct electrical measurement of the resistance between the base and the tip of a single cantilever of 20 Ω (corresponding to $R_S \sim 5$ Ω), a measurement of C $\sim 34$ pF, and a negligible loss tangent. For the radiofrequency amplitude and frequency listed above, the power dissipated in the trap is expected to be about 2 mW, or 0.5 mW per cantilever pair. As the geometry is scaled down in the chip plane (fixing $h$, $q$, and $V_0$ as above), we expect that the dissipated power per unit area of radiofrequency electrode should grow as $I_0 \propto s^{-2.2}$ (for aspect ratios $1 < s/h < 20$)[20].

In addition to stable trapping of individual ions in each of the two trapping zones, ions are shuttled between zones[30] by smoothly changing the voltages from trapping in one region to trapping in the adjacent region ~150 μm away. This has been demonstrated starting in either trap zone with shuttle times as fast as 2.5 ms, with the speed limited by low-pass filters installed on the chip.

Given these promising results for the GaAs microtrap architecture, we intend to fabricate different structures that will feature larger trap depths and may show lower heating rates by altering the electrode dimensions in the plane of the chip and increasing the separation between layers. We will also explore the fabrication of 'cross' and 'tee' junctions in the GaAs architecture for more advanced shuttling experiments, perhaps requiring a 3-layer geometry[25]. This symmetric high aspect ratio geometry could also accommodate other materials such as silicon, which may allow higher voltages to be applied with less radiofrequency dissipation. Ultimately, a hybrid geometry combining the symmetric high-aspect-ratio and asymmetric planar trap geometries might be considered. Here, the deeper symmetric cantilevered electrode zones might be used for loading and entangling zones where high trap strength and depth are required, and the planar trap zones might be used for complex shuttling operations.

## METHODS

**Fabrication.** The wafer (Fig. 1a) consists of a doped substrate on top of which are four layers grown by molecular beam epitaxy. Directly above the substrate is a 4 μm layer of $Al_{0.7}Ga_{0.3}As$, chosen for its insulating properties and selective etching versus GaAs. On top of it is a 2.3 μm layer of silicon-doped ($3\times10^{18}$ $e$/cm$^3$) GaAs, 4 μm of $Al_{0.7}Ga_{0.3}As$ and 2.3 μm of doped GaAs. As shown in Fig. 1, a series of dry and wet etch procedures define the cantilevered GaAs electrodes. The final step undercuts the $Al_{0.7}Ga_{0.3}As$ from the edges of the GaAs cantilever by about 15 μm to shield the trapped ion from the exposed insulator. Figure 2 shows a scanning electron micrograph of the final structure.

We attach the GaAs ion-trap chip to a ceramic chip carrier and attach 25-μm-diameter gold wires from the bond pads on the trap to the chip carrier, with a single wire connecting radiofrequency electrodes and individual wires going from the static-electrode bond pads to the chip carrier electrodes. The static electrodes are shunted to ground through 1,000 pF surface mount capacitors attached to the chip carrier, and our measurements show that the induced radiofrequency potential on the static electrodes is reduced to less than 1% of the applied radiofrequency potential. The chip carrier is then plugged into an ultra-high-vacuum-compatible socket that is permanently connected in the vacuum chamber. This arrangement allows for fast turnaround time; replacing an ion trap does not involve changing any other components inside the vacuum chamber.

**RF Delivery and Breakdown.** We apply radiofrequency potentials to the trap using a helical resonator with unloaded quality factor Q ≈ 500 and self-resonant frequency 54.9 MHz. When a capacitive coupler is impedance matched to the resonator-trap system, the resonant frequency falls to 15.9 MHz, and the unloaded quality factor of the system drops to 50. Breakdown of the AlGaAs layer appears to limit the amount of radiofrequency voltage that can be applied to the trap. We have applied a static potential as high as ~70 V between top and bottom cantilevers on a separate trap sample without breakdown, and a radiofrequency potential amplitude as high as $V_0 = 11$ V at 14.75 MHz before breakdown. We also observe nonlinear current-voltage behavior across the GaAs electrodes, where the measured current depends upon the polarity of the applied voltage and even the level of room lights at particular voltages. However, none of these effects were measurable at applied potentials below ~40 V and are thus not expected to play a role in the operation of the trap.

**Measurement of Heating using Raman Spectroscopy.** Heating of the secular harmonic motion of the trapped ion is measured by driving motion-sensitive stimulated Raman transitions between hyperfine ground states in the $^{111}Cd^+$ ion. A pair of laser beams each detuned ~70 GHz from the $^2S_{1/2} - {}^2P_{3/2}$ transition are directed onto the ion, with an optical beatnote near the 14.53 GHz atomic hyperfine splitting. The two Raman beams have a 7° angular separation, with the wavevector difference oriented 45° from the axis of the trap (axial Lamb-Dicke parameter of $\eta \approx 0.018$ for a trap frequency of 0.9 MHz). By adding varying delays $\tau$ after Doppler cooling but before the Raman probe, the increase of motional energy of the ion is reflected by the suppression of the Raman carrier transition rate through the Debye-Waller effect[27]. Assuming a thermal state of motion with average harmonic vibrational index $\bar{n}(\tau)$, the transition rate is proportional to $e^{-\eta^2\bar{n}(\tau)}$ in the Lamb-Dicke limit where $\eta^2\bar{n}(\tau) \ll 1$. Here we neglect the Debye-Waller factor from the more tightly-confined transverse motion, expected to be negligible compared with that of axial motion. After the delay $\tau$, the Raman transition rate is measured by interrogating the hyperfine level of the ion[3] after a time $t$ of exposure to the Raman probe, and fitting the initial development of the transition probability as a quadratic in time: $P(t) = \sin^2(Rt/2) \sim (Rt/2)^2$. We find that the Raman carrier transition rate $R$ decreases by approximately 25% after a delay of $\tau = 0.5$ ms (with a negligible effect of heating on the rate during the 10 μs Raman probe), as shown in Fig. 4a. This corresponds to an axial heating rate of $d\bar{n}(\tau)/d\tau = (1.0\pm0.5)\times10^6$ s$^{-1}$. The quoted error is dominated by the uncertainty in the absolute value of $\eta^2\bar{n}(\tau)$ that relates the Raman transition rate to the heating rate, in addition to the statistical uncertainty in the data.

**Acknowledgements** We acknowledge useful discussions with J. A. Rabchuk, S. Horst, T. Olver, K. Eng, P. Lee, P. Haljan, K.-A. Brickman, L. Deslauriers, and M. Acton. This work was supported by the U.S. Advanced Research and Development Activity and National Security Agency under Army Research Office contract W911NF-04-1-0234, and the National Science Foundation Information Technology Research Program.